\documentclass[twocolumn]{revtex4-1}
\usepackage[utf8]{inputenc}  
\usepackage[T1]{fontenc}  
\usepackage{helvet}
\usepackage{amsmath}
\usepackage{graphicx}
\usepackage{amssymb} 
\usepackage{url}
\usepackage{braket}
\newcommand{\SO}{\mathrm{SO}}
\newcommand{\SU}{\mathrm{SU}}

\begin{document}

\title{The spinorial ball (II): a manipulable qubit at human scale.}
\author{Samuel Bernard-Bernardet}
\affiliation{DotWave Lab, Chamb\'{e}ry, France}
\author{Benjamin Apffel}
\email{benjamin.apffel@epfl.ch}
\affiliation{Institute of Electrical and Micro Engineering, Laboratory of Wave Engineering, Ecole Polytechnique Federale de Lausanne (EPFL), Station 11, 1015 Lausanne, Switzerland}
\date{October 2024}

\begin{abstract}
The spinorial ball is an electronic manipulable device that we recently introduced to discuss the origin of spin-1/2 from rotations group representation, without relying on the quantum mechanics framework.  Nevertheless, it is also a macroscopic visualization of a quantum two-level system, and can thus be used to gain intuition on some generic features of qubits. The present article therefore aims to complement and extend our previous work by discussing how the spinorial ball can be used to visualize quantum mechanics features. The Bloch sphere, the Hopf fibration and the Berry phase can for instance easily be seen and manipulated using this original device. We also discuss how the spinorial ball can be used to visualize Hamiltonian evolution, and we describe an explicit mapping between the ball's motion and the evolution of 1/2-spin in arbitrary magnetic field. An electronic implementation of projective measurement that matches the predictions of quantum mechanics is also proposed. The present article is written as a practical guide to manipulate the ball and establishes the exact correspondence between the spinorial ball and a generic two-level system.
\end{abstract}
\maketitle
\section{Introduction}
\label{sec:intro}
Among all the remarkable features of quantum mechanics, spin-1/2 may be one of the must puzzling \cite{10.1119/1.11806}. It arises from representation of the rotation group in complex Hilbert space, of which it is difficult to build a clear mental image. To provide better insight, several works have proposed waves \cite{10.1119/1.14580,BELINFANTE1939887, deymier_2023,mahmood_2022} or mechanical \cite{leroy_simulating_2006,leroy_simulating_2010} analogs, but a representation in the direct space $\mathbb{R}^3$ was still missing. In a previous work, we have filled this gap and introduced the spinorial ball as a macroscopic object that exhibits spin-1/2 behavior under physical rotations \cite{bernard_2023}. The original goal was to discuss the mathematical roots of half-integer spin outside of quantum mechanics framework.

Nevertheless, the most striking physical observations related to spin-1/2 occur at the quantum scale. For instance, Stern-Gerlach experiment or anomal Zeeman effect can not be explained without half-integer spins \cite{basdevant_quantum_2005, cohen-tannoudji_quantum_1986,gerlach_experimentelle_1922}. It is therefore of prime interest to build intuition on this fundamental concept of quantum mechanics. Moreover, qubits are at the heart of quantum information \cite{Nielsen_Chuang_2010} which is currently an extremely active topic of research with strong potential applications. From this perspective, the existence of visualizing tools such as the spinorial ball can be extremely useful both for research and teaching purpose.

The present work therefore aims to explain how the spinorial ball can be practically manipulated to fully visualize a single qubit as well as several related mathematical structures. The paper is organized as follow. We first briefly summarize the main features of the spinorial ball in order to make this work as self-consistent as possible, the details being available in our previous publication \cite{bernard_2023}. We then discuss how to practically visualize the Bloch sphere, the Hopf fibration and topological phases by manipulating the ball. The last section discusses how to map any hermitian Hamiltonian evolution on the ball, as well as a procedure to perform projective measurement of a given state which statistics perfectly mimics quantum mechanics.

Note that the spinorial ball (which name comes from Ref. \cite{baez2014g2}) is more than anything an object that is made to be manipulated. We therefore invite the reader to build, if possible, his own spinorial ball using our open-source ressources \cite{gitSam}. Alternatively, a virtual version can be manipulated in an internet browser \cite{noauthor_httpproxy-informatiquefrquball_nodate} or by rotating a mobile phone \cite{spinPhone}.

\section{Visualization of a qubit}
\subsection{Main features of the spinorial ball}
\begin{figure*}
    \centering
    \includegraphics[width=17cm]{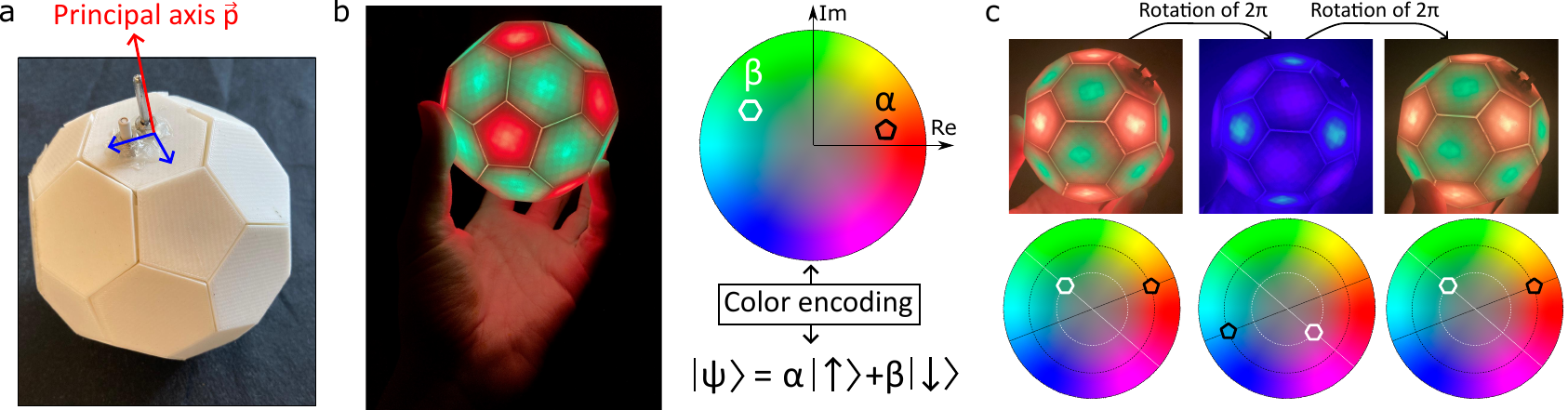}
    \caption{(a) Picture of the spinorial ball (b) Color encoding of complex number with modulus lower than one and several examples of qubit visualization $\ket \psi = \alpha \ket \uparrow + \beta \ket \downarrow$ with $|\alpha|^2+|\beta|^2 =1$. (c) Half integer spin behavior of the ball: starting from the left image with color state $\ket \psi$,  a full physical turn of the ball changes the state to $-\ket \psi$. If one performs a second turn, the color state is back to $\ket \psi$}
    \label{fig:1}
\end{figure*}
The spinorial ball consists in a truncated sphere made of pentagonal and hexagonal LED panels (see Fig. \ref{fig:1}a). Each type of panel displays a color that encodes a complex number in the unit disk as shown in Fig. \ref{fig:1}b. Two colors are therefore necessary and sufficient to encode any vector state of a two-level system of the form
\begin{equation}
    \ket \psi = \alpha \ket \uparrow + \beta \ket \downarrow, \quad (\alpha, \beta) \in \mathbb{C}^2 / |\alpha|^2 + |\beta|^2 = 1
    \label{eq:generalState}
\end{equation}
where $\ket \uparrow$ and $\ket \downarrow$ are two orthogonal vector forming a basis of the Hilbert space, and are chosen as eigenstates of $\sigma_z$ in what follows. The colors corresponding to $\alpha$ and $\beta$ will be displayed on all the pentagons and all the hexagons respectively, an example of such encoding being shown in Fig.\ref{fig:1}b.

When the ball is turned on, it is initialized in the state $\ket{\psi_0} = \ket \uparrow$ corresponding to red pentagons and black hexagons. The color displawed by the ball are changed upon rotations, which are encoded in elements of SO(3) and measured practically using a gyroscope and an Arduino placed in the ball. To keep track of the ball's orientation during the motion, one can imagine that a direct trihedron is fixed on it as in Fig. \ref{fig:1}, with one of the axis  $\vec p$ pointing toward $\vec z$ when the ball is initially in $\ket \uparrow$. This axis $\vec p$ will be called in what follows the principal axis and will be of interest in the following.

In order to explain how the colors are changed upon rotations, we parameterize elements of SO(3) as
\begin{equation}
    R_{\vec n} (\theta) = e^{-i \theta \vec J . \vec n}
    \label{eq:SO3}
\end{equation}
with $\theta \in [0, \pi[$, $\vec n$ a unit vector and $-i \vec J =-i (J_x, J_y, J_z)$ containing the three generators of SO(3). One element of SO(3) uniquely defines an orientation of the ball (and of the fictious trihedron glued on it). On the other hand, the unitary operators of SU(2) acting on spinors (\ref{eq:generalState}) can be parametrized in a similar manner \cite{appel_mathematics_2007}
\begin{equation}
    S_{\vec n} (\theta) = e^{-i \theta \vec \sigma . \vec n/2}
    \label{eq:SU2}
\end{equation}
with $\vec \sigma = (\sigma_x, \sigma_y, \sigma_z)$ the Pauli matrices. The operator $V: S_{\vec n} (\theta) \in \SU(2) \longrightarrow R_{\vec n} (\theta) \in \SO(3)$ that maps Eq. (\ref{eq:SU2}) to Eq. (\ref{eq:SO3}) is a group homomorphism \cite{appel_mathematics_2007}, and establishes the correspondence between the ball's rotations and the displayed colors. \footnote{It is very tempting to absorb the $1/2$ factor of Eq. (\ref{eq:SU2}) either in $\theta$ or in $\vec \sigma$ to recover full symetry with Eq. (\ref{eq:SO3}). However, doing the first choice would break the group homomorphism property of $V$ while the second one would prevent $\vec J$ and $\vec \sigma /2$ to verify the same commutations relations. This factor $1/2$ must therefore remain, which will have crucial consequences.}

All the successive orientation of the ball between $t=0$ and $t=T$ are described by a continuous family of rotations $\{ R(t) \}_{0 \leq t \leq T}$, which is a path in SO(3) that goes from $R(0)=\mathbb{I}_3$ (initial orientation) to $R(T)$ (final orientation). This path in SO(3) can be  uniquely lifted in SU(2) through $V$ as a continuous path $\{ S(t) \}_{0 \leq t \leq T}$ that goes from $S(0)=\mathbb{I}_2$ to $S(T)$. This uniquely defines the action on the color state that will go from $\ket{\psi_0}$ to $S(T) \ket{\psi_0}$ as the ball is rotated. Any unitary operator can therefore be applied to the displayed spinor by performing the suitable rotation of the physical ball.

We show in Figure \ref{fig:1}c an example of the ball's manipulation. Starting from a given state, we perform a $2\pi$ rotation around any vector $\vec n$. The physical ball is left invariant by such transformation, but the complex numbers encoded in the colors have been changed to their opposite (Fig. \ref{fig:1}c). This is consistent with the intuitive understanding of spin 1/2 that goes to its opposite  $\ket{\psi_0} \rightarrow -\ket{\psi_0}$  after a full turn. It can also be seen in formula (\ref{eq:SU2}) and (\ref{eq:SO3}), in which taking $\theta=2\pi$ gives $+\mathbb{I}_3 \in SO(3)$ but $-\mathbb{I}_2 \in SU(2)$. When another $2\pi$ rotation is performed, the colors are back to their initial value, thus further confirming the 1/2-spin behavior. The fact that $2\pi$ and $4\pi$ rotations are not equivalent for complex spinors is related to the existence of two homotopy classes of loops in SO(3), which is discussed in details in our previous work \cite{bernard_2023}.

By design, the spinorial ball acts as a macoscropic qubit, which orientation is encoded through the colors in two complex components. It provides in particular an original visualization of quantum superposition. We now propose to describe how several important features of qubits can be seen on the ball by applying peculiar rotations.

\subsection{Seing the Bloch sphere with the ball}

\begin{figure*}
    \centering
    \includegraphics[width=17cm]{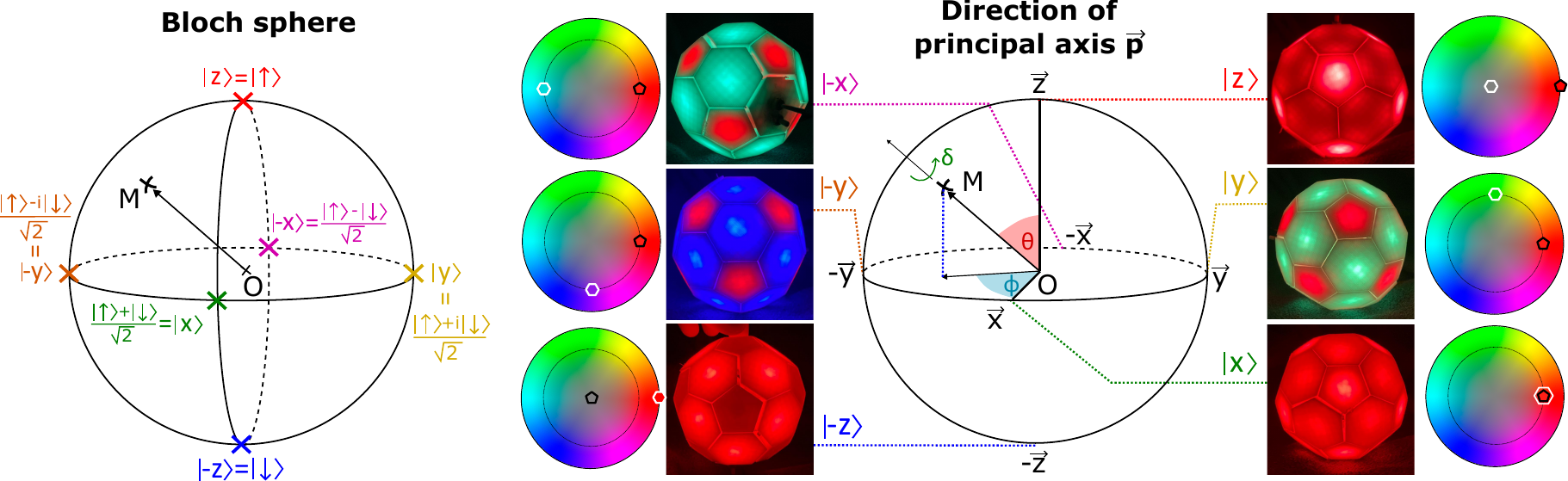}
    \caption{(left) The Bloch sphere for which each vector state corresponds to a point of the unitary sphere. (right) A state $M$ of the Bloch sphere is obtained, starting from $\ket \uparrow$ and principal axis along $\vec z$, by orienting the principal axis $\vec p$ of the spinorial ball along $\vec{OM}$. Remarkable points of the Bloch sphere are shown on the ball, the global phase having been chosen to keep the pentagons red (i.e. real positive).}
    \label{fig:2}
\end{figure*}

Since global phase factor is irrelevant in quantum mechanics, the vectors states in Eq. (\ref{eq:generalState}) are often parametrized as 
\begin{equation}
    \ket \psi = \cos{\left(\frac{\theta}{2}\right)} \ket \uparrow + e^{i\phi} \sin{\left(\frac{\theta}{2}\right)} \ket \downarrow,  (\theta, \phi) \in [0, \pi] \times [0, 2\pi[
    \label{eq:PsiBloch}
\end{equation}
Such state $\ket \psi$ can be mapped on a point $M$ on the unit sphere $S^2$ of $\mathbb{R}^3$ with angular coordinates $(\theta, \phi)$ 
\begin{equation}
    \vec{OM} =\cos(\phi) \sin(\theta) \vec x + \sin{(\phi)} \sin{(\theta)} \vec y + cos{(\theta)} \vec z
    \label{eq:OM}
\end{equation}
This geometrical representation of qubit states is the Bloch sphere and is depicted in Fig. \ref{fig:2}. The spinorial ball offers a convenient way to visualize it. We initialize it in state $\ket \uparrow$ with its principal axis $\vec p$ pointing toward $\vec z$. If we rotate the ball so that its principal axis $\vec p$ now points toward $\vec{OM}$ parametrized by Eq. (\ref{eq:OM}), then the displayed state corresponds (up to global phase) to the state $M$ of Eq. (\ref{eq:PsiBloch}). In other words, the state of the Bloch sphere displayed by the colors is consistent with the direction pointed by the principal axis of the ball, as illustrated in Fig. \ref{fig:2}. For instance, when the principal axis points toward $-\vec z$, the displayed state is $\ket \downarrow = \ket{-z}$. When it points toward $\vec x$ (resp. $-\vec x$), the color state is, up to global phase, $\ket x = \frac{\ket \uparrow + \ket \downarrow}{\sqrt{2}}$ (resp. $\ket{-x}=\frac{\ket \uparrow - \ket \downarrow}{\sqrt{2}}$). 

Describing a quantum state with the Bloch sphere uses the fact that the global phase of $\ket \psi$ is irrelevant in quantum mechanics. In particular, $\ket \psi$ and $-\ket \psi$ cannot be distinguished on the Bloch sphere since they are represented by the same point. On the other hand, they can be distinguished on the spinorial ball as shown in Fig. \ref{fig:1}c. This shows that the spinorial ball contains more information than the Bloch sphere, which corresponds to the Hopf bundle \cite{hopf_1931}.

\subsection{Hopf bundle}

\begin{figure*}
    \centering
    \includegraphics[width=17cm]{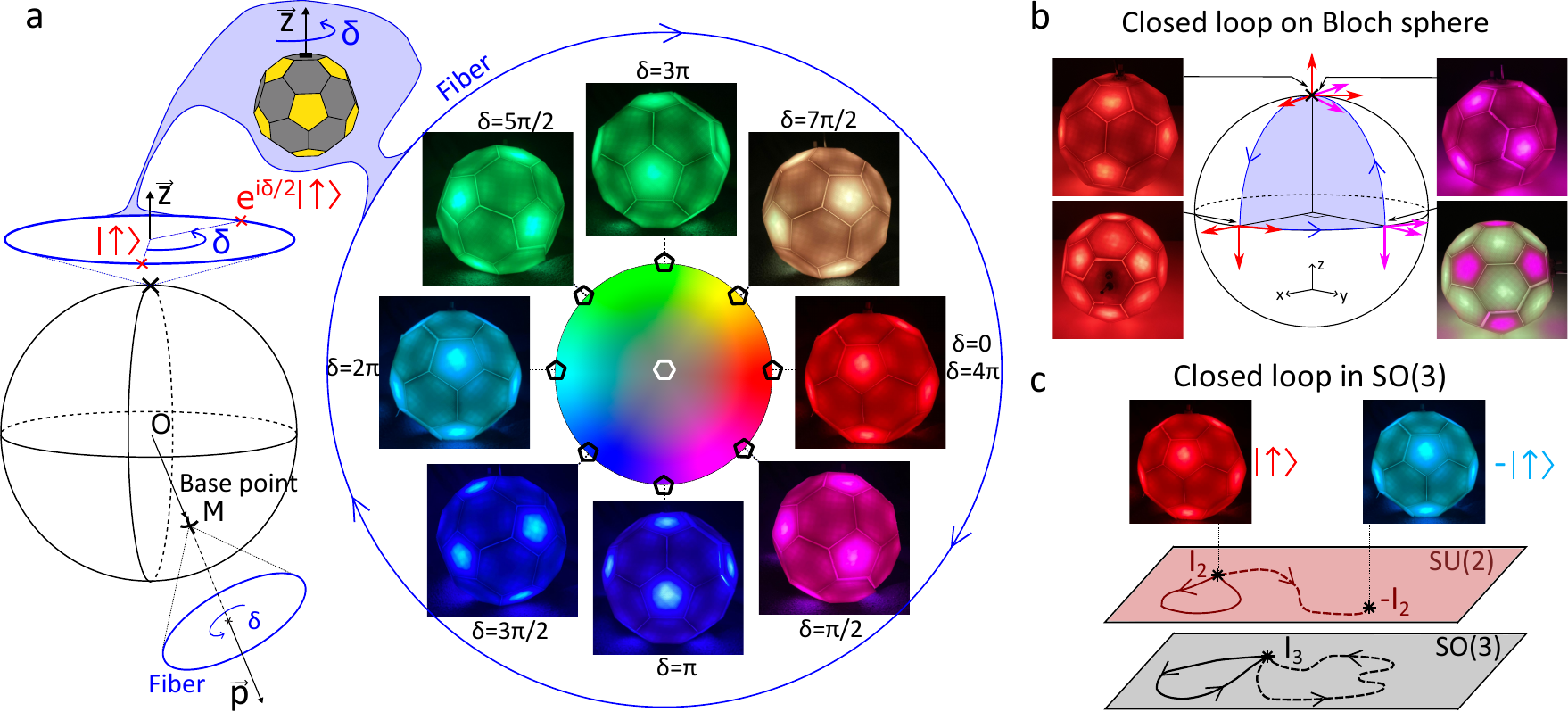}
    \caption{(a) Hopf bundle: a circle above each point $M$ of the Bloch sphere encodes the global phase. One can travel along each circle by rotating the ball around its principal axis axis $\vec{OM}$, which leaves the latter unchanged. An illustration of the fiber along $\vec z$ spanned by rotating the ball initially in $\ket \uparrow$ around $\vec z$ is shown aside. (b) Example of Berry phase introduced by a closed loop on the Bloch sphere. The phase difference between the initial and the final state ($-\pi /4$) is minus half of solid angle enclosed by the path ($\pi/2$). (c) $0$ or $\pi$-phase obtained by performing closed loop in SO(3), depending on whether the lift of the loop in SU(2) is a closed loop or an open path.}
    \label{fig:25}
\end{figure*}

To clearly demonstrate the impact of global phase, we first consider the case where the ball is initialized in $\ket \uparrow$ with $\vec p$ aligned with $\vec z$ as in Figure \ref{fig:25}. Performing a rotation of $\delta$ along $\vec z$ gives the vector state $e^{-i\delta/2} \ket \uparrow$ (Eq. (\ref{eq:SO3}-\ref{eq:SU2})). The latter is identified with $\ket \uparrow$ on the Bloch sphere, as both states only differ by a global phase, but do not correspond to the same color illumination of the ball. More generally, each state $\ket \psi$ on the Bloch sphere can be thought as representing the whole set of states $\{e^{i\delta} \ket \psi, 0 \leq \delta \leq 2\pi \}$. One can therefore associate to each point of the Bloch sphere a unit circle denoted $S^1$ to encode the global phase $\delta$ as in Fig. \ref{fig:25}. Mathematically speaking, those circles are called fibers while the Bloch sphere is called the base space, and the total structure is the Hopf bundle.

The spinorial ball allows to distinguish two states differing by a global phase and thus to visualize the Hopf bundle. For a given orientation with Euler angles $(\theta, \phi, \delta)$ corresponding to nutation, precession and intrisic rotation respectively \footnote{Note that this convention is not the same as in Fig. 2 for the angle $\phi$. The convention used for Eq. (6) corresponds to the sequence of rotations (1) Change the latitude by $\theta$ (2) Change the longitude by $\phi$ and (3) Perform rotation around proper axis $\vec p$ by $\delta$. In particular, this convention implies $\phi = 0$ after the first step, while it could be non-zero in the convention of Fig. 2.}, the two possible spinor displayed by the ball are (taking $\theta \neq 0 [\pi]$)
\begin{equation}
    \pm \ket \psi = e^{-i\frac{\delta+\phi}{2}} \cos{\left(\frac{\theta}{2}\right)} \ket \uparrow + e^{-i\frac{\delta-\phi}{2}} \sin{\left(\frac{\theta}{2}\right)} \ket \downarrow.
\end{equation}
The choice of the sign is determined as before by the past history of the ball's motion. A rotation by an angle $\delta$ around $\vec{OM}$ corresponds to the transformation $\ket \psi \rightarrow e^{-i \delta/2}\ket \psi$ on the color state, which corresponds to travel along the fiber attached to the base point $M$ on the Bloch sphere. As the ball's principal axis $\vec{OM}$ is left invariant by such rotation, the corresponding state on the Bloch sphere can still be found by looking at the direction of $\vec p$. Both hexagon's and pentagon's colors vary during the rotation, but their modulus as well their phase difference (if defined) remains constant along the rotation. This gives moreover a geometrical picture of Hopf bundle: the base state corresponds to the direction indicated by the principal axis of the ball, while the fiber corresponds to the last degree of freedom for the ball's rotation around its principal axis once the later is fixed. The figure \ref{fig:25} shows an example with the ball initially in the state $\ket \uparrow$ rotated around $\vec z$, and shows the different states accessible in the fiber of $\ket \uparrow$. Considering all the Bloch sphere's points and all the fibers, one recovers all the possible phase states of the form of Eq. (\ref{eq:generalState}). This shows that the unit sphere of $\mathbb{C}^2$ containing all the vector states (\ref{eq:generalState}) can locally be decomposed as $U\times S^1$, with $U \subset S^2$ an open subset of the Bloch sphere and $S^1$ the unit circle. However, such decomposition cannot be made globally, which makes the Hopf bundle a non-trivial fiber bundle (see for instance \cite{10.21468/SciPostPhysLectNotes.39} for details).

\subsection{Berry phase and homotopic phase}

As the Hopf fibration is accessible, the geometrical Berry phase  \cite{pancharatnam_1956,berry_1984} can also be visualized on the ball. To do so, one can for instance move the principal axis of the ball along the blue path of Fig. \ref{fig:25}b consisting of three rotations of $\pi/2$, the two other vectors of the trihedron being parallely transported. The principal axis points toward $\vec z$ both before and after the sequence of motions, so that the Bloch sphere state is $\ket \uparrow$ in both cases. However, the state represented on the ball went from $\ket \uparrow$ to $e^{-i\pi/4} \ket \uparrow$, showing that performing the closed loop on the Bloch sphere has led to the emergence of an extra phase of $\phi_B=-\pi/4$, which is an instance of the celebrated Berry phase. This experiment on the spinorial ball is directly analogous to study a spin in $\ket \uparrow$ state subjected to magnetic field which direction slowly varies along time \cite{doi:10.1126/science.1149858}. In the latter case, the Berry phase is minus half the solid angle enclosed by the closed path drawn by the magnetic field on the Bloch sphere \cite{Bernevig+2013}. This is consistent with the previous result, as the solid angle enclosed by the ball's motion is $\pi/2$, and also matches up to sign the $\pi/4$ angle difference between the tangent vectors before and after parallel transport  (see Fig. \ref{fig:25}b). Performing the same experiment with the $\ket \downarrow$ state gives opposite sign $\phi_B = \pi/4$, and can also be visualized with the ball. Of course, the link between the Berry phase and the enclosed solid angle only holds if no rotation is performed around the principal axis $\vec{OM}$ when $M$ follows geodesic (i.e. great circles) on the sphere. This condition ensures the parallel transport of the two axis orthogonal to $\vec p$ as in Fig. \ref{fig:25}b. If one wants to visualize the Berry phase induced by following, say, the 45 degree parallel, one should add a gradual rotation along the principal axis while the path is followed to ensure parallel transport of the trihedron.

Another type of topological phase, related to the two homotopy classes of paths of SO(3) \cite{appel_mathematics_2007}, can also be observed on the ball in the following way. Starting from $\ket \uparrow$, we perform a sequence of rotations such that the initial and final trihedron are the same, thus defining a closed loop $\mathbb{I}_3 \rightarrow \mathbb{I}_3$ in SO(3) (Fig. \ref{fig:25}c). Depending whether this loop is contractible or not, it can either correspond to a closed loop $\mathbb{I}_2 \rightarrow \mathbb{I}_2$ or to an open arc $\mathbb{I}_2 \rightarrow -\mathbb{I}_2$ in SU(2) (Fig. \ref{fig:25}c). As a consequence, the phase difference $\gamma$ between the final and the initial state of the ball is either $0 [2\pi]$ or $\pi [2\pi]$. For instance, a rotation of $2\pi$ around any axis leads to $\gamma = \pi$, while a rotation of $4\pi$ gives $\gamma = 2\pi$. An intuitive way to think about it is to imagine that the spinorial ball moves along the three dimensional analog of a Moebius strip.  

The value of $\gamma$ only depends on the homotopy class of the chosen loop, and is therefore insensitive to continuous deformations of the latter. This is in strong contrast with the Berry phase $\phi_B$ discussed above, which depends on the path's geometrical details. Note also that rotating the $\ket \uparrow$ state by $2\pi$ around the $\vec z$ axis leads to $\gamma = \pi$, while $\phi_B =0$ since the solid angle enclosed by the corresponding path on the Bloch sphere is zero. This emphasize the different origin of the two phases, which result respectively from closed loops in SO(3) and parallel transport on the sphere. A closed loop in SO(3) imposes the whole orientation of the ball to be invariant,  while a closed loop on the Bloch sphere only imposes the principal axis to be left unchanged. This added restriction for SO(3) loops constrains the homotopic phase $\gamma$ to be a multiple of $\pi$. On the other hand, the Berry phase $\phi_B$ can take any values depending on the closed loop performed on the Bloch sphere. 

The spinorial ball hence offers striking illustrations of Hopf fibration and of the associated topological phases, which can be measured by looking at the color difference before and after a sequence of motions. Contrary to quantum mechanics, there is no need to define a Hamiltonian nor any adiabatic condition to observe them, which emphasizes their geometrical nature. A Hamiltonian evolution can nevertheless be mapped on the ball, as discussed in the next section.

\section{Mapping quantum dynamics on the ball}
The most general evolution of a single qubit can be decomposed as a sequence of Hamiltonian evolutions followed by projective quantum measurements. We show in this last section how both types of evolution can be implemented on the spinorial ball, which therefore allows to visualize all the possible quantum evolution of a single qubit.

\subsection{Hamiltonian evolution of the ball}

Let us assume that the ball is currently in the state $\ket{\psi(t)} = S(t) \ket{\psi_0}$ and that a small rotation of angle $\delta(t) \ll 1$ around the axis $\vec n(t)$ occurs between $t$ and $t + dt$. The associated spin state is now $\ket{\psi(t+dt)} = S_{\vec n} (\delta) \ket{\psi(t)} \approx (\mathbb{I} - i \delta \vec n . \vec \sigma  /2) \ket{\psi(t)}$. Taking $dt \rightarrow 0$ leads to the evolution equation
\begin{equation}
    \partial_t \ket \psi = -i  \vec \Omega(t) . \vec \sigma \ket{\psi} = - i H(t) \ket \psi
\end{equation}
where $\vec \Omega(t) = \frac{1}{2} \frac{d(\delta \vec n)}{dt}$ encodes the instantaneous rotation. The previous relation shows that any Hamiltonian evolutions $H(t)$ of zero trace can be observed on the ball providing one applies the suitable rotation on it. In particular, the evolution of a qubit in any magnetic field $\vec B(t)$ described by $H = - \vec B(t) . \vec \sigma$ can be implemented by identifying $\vec \Omega(t) = \vec B(t)$ and taking the Planck constant $\hbar = 1$. The Larmor precession in constant magnetic field $B_0 \vec z$ can for instance be implemented by performing a rotation around $\vec z$ at rate $B_0$. 

Note that the zero trace property of the Hamiltonian mentioned above is not a strong constrain, as a non-zero trace can always be absorbed in the global phase of $\ket \psi$. Any hermitian Hamiltonian evolution, including most of quantum gates, can therefore be implemented on the ball. We now discuss how quantum measurement can be implemented electronically on the ball.

\subsection{Quantum measurement}

\begin{figure}
    \centering
    \includegraphics[width=8cm]{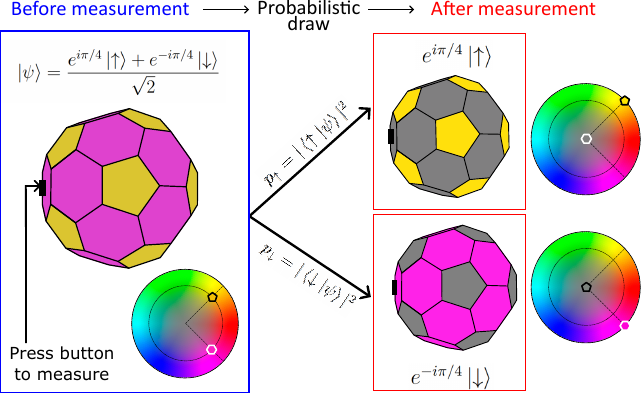}
    \caption{(a)Example of measurement simulation: starting from $\ket \psi = (e^{i\pi/4}\ket \uparrow + e^{-i\pi/4}\ket \downarrow)/\sqrt{2}$, pressing the button reinitialize the ball in $e^{i\pi/4} \ket \uparrow$ or in $-e^{i\pi/4} \ket \downarrow$ with probability $|\braket{\uparrow|\psi}|^2$ and $|\braket{\downarrow|\psi}|^2$ respectively. }
    \label{fig:3}
\end{figure}

In quantum mechanics, one can performs a spin measurement on the state $\ket \psi =  \alpha \ket \uparrow +  \beta \ket \downarrow$ along the $z$-axis using for instance a Stern-Gerlach experiment. The results can only take two values $\hbar/2$ or $-\hbar/2$ with respective probabilities $p_\uparrow = |\braket{\uparrow|\psi}|^2=|\alpha|^2$ and  $p_\downarrow = |\braket{\downarrow|\psi}|^2=|\beta|^2$. Depending on the result, the spin state after measurement is projected on $\ket \uparrow$ or $\ket \downarrow$. Such measurement process can be mimicked on the spinorial ball using an external button which commands, when pushed, that a measurement must be performed. In such case, the probability $p_\uparrow=|\braket{\uparrow|\psi}|^2$ is computed on the current state and a random number $t$ is uniformly picked between $0$ and $1$. If $t < p$ (resp. $t \geq p$), the display on the LED panel is reinitialized in the normalized state $\frac{\braket{\uparrow|\psi}}{\sqrt{p}} \ket \uparrow$ (resp. $\frac{\braket{\downarrow|\psi}}{\sqrt{1-p}}\ket \downarrow$) as depicted in Fig. \ref{fig:3}. Doing so, the spinorial ball reproduces the corresponding wave function collapse and the the result's statistics over many realization will exactly match the one expected from quantum mechanics. 

As in quantum mechanics, measurement strongly affects the spin state and the color state $(\alpha, \beta)$ undergoes some discontinuity during measurement. Contrary to previous sections where a rotation could always be inverted by performing the opposite motion, the measurement process is not invertible and the initial state is irremediably lost. There is however a strong difference with quantum mechanics:  the complex state $\ket \psi$, and thus the probability distribution, can be directly observed on the ball. Nevertheless, it is not enough to predict the result of a single measurement, except for degenerate cases $\ket \psi = \ket \uparrow$ or $\ket \downarrow$ where the draw is deterministic. In all other cases, individual measurement is still a random process, and only its average over many realization will match the expectation from the color's intensity.

So far, we have only described how to perform measurement along the $\vec z$ axis, as the natural basis to perform measurement is  $(\ket{\uparrow}, \ket{\downarrow})$ due to the way pentagon's and hexagon's colors are chosen. If we now want to perform measurement along an arbitrary axis $\ket n = S \ket \uparrow$ and $\ket{-n} = S \ket \downarrow$, we can compute the probabilities $p_{+} = |\braket{n|\psi}|^2 = |\braket{\uparrow| S^\dagger \psi}|^2$ and $p_{-} = |\braket{-n|\psi}|^2 = |\braket{\downarrow| S^\dagger \psi}|^2$. We immediately see that performing measurement on $\ket \psi$ along $\vec n$ is equivalent to perform measurement on $S^\dagger \ket \psi$ along $\vec z$, which simply states the equivalence between rotating the detector by $S$ or rotating the ball by $S^\dagger=S^{-1}$. Applying suitable rotations on the ball before measurement thus allows to perform measurement of $\ket \psi$ along any axis. The full procedure works as follow. First, one must rotate the ball in order to apply $S^\dagger$ to $\ket \psi$. One can then press the button, which performs measurement along $\vec z$ and projects $\ket \psi$ on the $\ket \uparrow$ or $\ket \downarrow$ state. Last, one must 'derotate' the ball, which corresponds to apply $S$ to the projected state. After the whole process, the ball will be in state $\ket n$ with probability $p_+$ and in state $\ket{-n}$ with probability $p_-$, in line with what is expected.

\section{Conclusion and perspectives}
This manuscript, along with our previous work, provides a comprehensive description of the spinorial ball and the associated concepts that can be visualized on it. These include half-integer spin, quantum superposition, the Bloch sphere and Hopf bundle, the Berry phase, and quantum evolution. It also introduces several advanced mathematical concepts, such as path lift, path homotopy classes, and fiber bundles. The spinorial ball can be used in classrooms to support quantum mechanics lectures or for visualization purposes. Specifically, the effects of quantum logic gates on qubits, such as Hadamard's or Pauli's gates, can be illustrated. Its design and corresponding codes are freely available on our GitHub. Our work also suggests developing similar visualization tools for higher spin values, which could be achieved by using a greater number of colors
\section{Acknowledgment}
Parts of this project were developed during ``Les Gustins'' Summer School with support of Jean Baud and Ingénieurs et Scientifiques de France - Sillon Alpin (IESF-SA). The authors thank these organizations and acknowledge attendees for stimulating discussions. The authors have no conflicts of interest to disclose.
\bibliographystyle{unsrt}
\bibliography{biblio}
\end{document}